\documentclass{aa}

\usepackage{graphicx,natbib}
\usepackage{bm}
\usepackage{amsmath}

\usepackage{color}
\usepackage{ulem}

\begin{document}

%
   \title{Temperature dependent growth rates of the upper-hybrid waves and solar radio zebra patterns}

   \authorrunning{Ben\'a\v{c}ek et al.}
   \titlerunning{The growth rate of upper-hybrid waves}

   \author{J. Ben\'a\v{c}ek$^{1}$, M. Karlick\'y$^2$ and L.V. Yasnov$^3$
          }
   \offprints{, \email{jbenacek@physics.muni.cz}}

   \institute{Department of Theoretical Physics and Astrophysics, Masaryk University,
   Kotl\'a\v{r}sk\'a 2, CZ -- 611 37 Brno, Czech Republic
         \and
   Astronomical Institute of the Czech Academy of Sciences, Fri\v{c}ova 258, CZ -- 251 65 Ond\v rejov, Czech Republic
         \and
         St.-Petersburg State
                  University, St.-Petersburg, 198504, Russia\\}
   \date{Received ; accepted }


  \abstract
   {The zebra patterns observed in solar radio emission are
   very important for flare plasma diagnostics. The most promising model of these patterns
   is based on double plasma resonance instability, which generates upper-hybrid waves, which can be then transformed into the zebra emission.}
   {We aim to study in detail the double plasma resonance instability of hot
   electrons, together with a much denser thermal background plasma.
In particular, we analyse how the growth rate of the instability depends on the temperature of both the hot
   plasma and background plasma components.}
   {We numerically integrated the analysed model equations, using Python and Wolfram Mathematica.}
   {We found that the growth-rate maxima of the upper-hybrid waves for non-zero temperatures of both
the   hot and background plasma are shifted towards lower frequencies comparing
   to the zero temperature case. This shift increases with an increase of
   the harmonic number $s$ of the electron cyclotron frequency and temperatures of both hot and background plasma components.
   We show how this shift changes values of the magnetic field strength estimated from observed zebras.
   We confirmed that for a relatively low hot electron temperature, the dependence of growth rate vs. both the ratio
   of the electron plasma and electron cyclotron frequencies expresses distinct
   peaks, and by increasing this temperature these peaks become smoothed.
   We found that in some cases, the values of wave number vector components for the upper-hybrid wave
   for the maximal growth rate strongly deviate from their analytical
   estimations. We confirmed the validity of the assumptions used when deriving model equations.}
   {}

   \keywords{Sun: radio radiation -- Instabilities -- Methods: analytical}

   \maketitle

\section{Introduction}

Zebra patterns (zebras) are fine  IV radio-burst type 
structures, observed during solar flares in the dm- and m-wavelength ranges
\citep{1972SoPh...25..210S, 2012A&A...538A..53C}. They are considered to be an
important source of information about the plasma density and intensity of the
magnetic field in their sources.

There are many zebra models \citep{1972SoPh...25..188R,
1975PhDT.........1K, 1975SoPh...44..461Z, 1976SvA....20..449C,
1990SoPh..130...75C, 2003ApJ...593.1195L, 2007SoPh..241..127K,
2006A&A...450..359B, 2006SoPh..233..129L, 2009PlPhR..35..160L,
2010Ap&SS.325..251T, 2013A&A...552A..90K}. The most promising one is based on
the double plasma resonance instability of the plasma, together with
loss-cone type electron distribution function \citep{1975SoPh...44..461Z,
2013SoPh..284..579Z}. In this model, the upper-hybrid waves are
 first generated, then transformed to electromagnetic (radio) waves
with the same (fundamental branch) or double frequency (harmonic branch) as the
upper-hybrid waves. This model has with simple resonance condition
\begin{equation}
\omega_\mathrm{UH} \approx s \omega_\mathrm{B},
\label{eq00}
\end{equation}
where  $\omega_\mathrm{UH} = \sqrt{\omega_\mathrm{p}^2  + \omega_\mathrm{B}^2}$
is the upper-hybrid frequency of the background plasma, $\omega_\mathrm{p}$ and
$\omega_\mathrm{B}$ are electron-plasma and electron-cyclotron frequencies and
$s$ is a integer harmonic number, is used for estimations of the magnetic field
strength and electron plasma density in zebra radio sources
\citep{2001SoPh..202...71L,2013SoPh..284..579Z,2015A&A...581A.115K}. However,
this resonance condition is only valid in the zero-temperature limit. If we
want to analyse effects of temperatures on the zebra generation processes, we need
to take the resonance condition in its general form, see relation
\ref{eq13}.

In this paper we study these temperature effects in detail. It is
shown that these effects require a correction in the method used to estimate magnetic field strength in zebra radio sources, especially for zebra
stripes at high harmonics.

The paper is structured as follows: In Section 2 we present model equations
describing the double plasma resonance instability and methods of their
solution. Results are summarized in Section 3. Finally, the paper is completed
by discussions and conclusions in Sections 4 and 5.

\section{Model}

Similarly to \citet{1986ApJ...307..808W, 2004SoPh..219..289Y},
for the hot electron component we consider the DGH distribution function with
the parameter $j=1~$\citep{1965PhRvL..14..131D}
\begin{equation}
f = \frac{u_\perp^2}{2 (2\pi)^{3/2} v_\mathrm{t}^5} \exp \left(-\frac{u_\perp^2 + u_\parallel^2}{2 v_\mathrm{t}^2}\right),
\label{edory}
\end{equation}
where $u_\perp = p_\perp/m_\mathrm{e}$ and $u_\parallel =
p_\parallel/m_\mathrm{e}$ are electron velocities and $p_\perp$ and
$p_\parallel$ are components of the electron momentum perpendicular and
parallel to the magnetic field, and $m_\mathrm{e}$ is the electron mass. For
simplification and in agreement with \citet{1986ApJ...307..808W}, we call
$v_\mathrm{t}$ here the thermal velocity of hot electrons and we use the
term the temperature of hot electrons, although the distribution function in
relation~\ref{edory} is not Maxwellian.

Unlike \citet{1986ApJ...307..808W}, we included the
background plasma with non-zero temperature. Its density $n_\mathrm{b}$ is
assumed to be much greater than the density of the hot electrons
$n_\mathrm{h}$.

In agreement with the approach by \citet{1982ApJ...259..844M,
1986ApJ...307..808W, 2004SoPh..219..289Y}, we use the condition for double
plasma resonance instability in the form
\begin{equation}
    \omega_\mathrm{UH} -\frac{k_\parallel u_\parallel}{\gamma} - \frac{s \omega_\mathrm{B}}{\gamma} = 0,
\label{eq13}
\end{equation}
where  $\omega_\mathrm{UH} = \sqrt{\omega_\mathrm{p}^2  + \omega_\mathrm{B}^2 +
3 k^2 v_\mathrm{tb}^2}$ is the upper-hybrid frequency of the background plasma,
where the temperature effect is included, $v_\mathrm{tb}$ is the thermal
velocity of the background plasma, $\gamma = \sqrt{1 + u^2 / c^2}$ is the
Lorentz factor, $k$ is the absolute value of the wave number vector. Its componets
 $k_\parallel$ and $k_\perp$ are in parallel and perpendicular
directions to the magnetic field.

Then, starting from the basic equations presented in \citet{1986ApJ...307..808W,
2004SoPh..219..289Y}, we derived the relation for the growth rate
$\Gamma_\mathrm{UH}$ of the upper-hybrid waves as
\begin{eqnarray}\label{eq1}
    \frac{\Gamma_\mathrm{UH}}{\omega_\mathrm{B} n_\mathrm{h}} &=& \frac{\pi^{2} e^{2}}{\omega_\mathrm{B}^{2} m_\mathrm{e} \sqrt{r_\mathrm{pB}^{2} + 1} } \sum_s  \int_0^{u_{\perp,\mathrm{max}}} (h^+ + h^-) du_{\perp}, \\
    \label{eq2}
    h^{\pm} &=& \frac{s^{2} G^{\pm} e^{-\frac{1}{2 v_\mathrm{t}^2} \left( u_\parallel^{\pm 2} + u_\perp^{2} \right)
} J^{2}_{s}\left(\frac{\sqrt{\lambda} u_\perp}{v_\mathrm{t}}\right)}{2\sqrt{2\pi} \lambda  f^{\pm} v_\mathrm{t}^{5} \left(u_\parallel^{\pm 2} + u_\perp^{2} + 1\right)}, \\
    \label{eq3}
    f^{\pm} &=& \frac{s u_\parallel^{\pm} - \frac{u_\perp^{2} + 1}{\sigma v_\mathrm{t}}}{\left(u_\parallel^{\pm2} + u_\perp^{2} + 1\right)^{3/2}}, \\
    \label{eq4}
    u_{\parallel}^{\pm} &=&
\frac{1}{\beta} \left(\sigma s v_\mathrm{t} \mp \sqrt{\beta + 1} \sqrt{- \beta \left(u_\perp^{2
} + 1\right) + \sigma^{2} s^{2} v_\mathrm{t}^{2}}\right),\\
    \label{eq5}
    \beta &=& -1 + v_\mathrm{t}^2 \sigma^2 (1+r_\mathrm{pB}^2) + 3v_\mathrm{tb}^2(1+\lambda \sigma^2),
\end{eqnarray}
where $n_\mathrm{h}$ is the hot electron density, $e$ is the electron
charge, $r_\mathrm{pB}$ is the ratio $\omega_\mathrm{p}/\omega_\mathrm{B}$,
$J_s(x)$ is the Bessel function of first kind of $s$-th order and $\lambda,
\sigma$ are dimensionless parameters
\begin{equation}\label{eq6}
    k_\perp = \sqrt{\lambda} \frac{\omega_\mathrm{B}}{v_\mathrm{t}}, \qquad k_\parallel = \frac{\omega_\mathrm{B}}{\sigma v_\mathrm{t}},
\end{equation}
and $G^{\pm}$ are functions
\begin{eqnarray} \label{eq6a}
G^{\pm} &=& g_1 + g_2^{\pm}, \\
g_1 &=& s \left(- u_\perp^{3} + 2 u_\perp v_\mathrm{t}^{2}\right), \\
g_2^{\pm} &=& \frac{u_{\perp}^3 u_{\parallel}^{\pm}}{ s v_t\sqrt{u_\parallel^{\pm2} + u_\perp^{2} + 1 }  }.
\end{eqnarray}
The term $g_2^{\pm}$ comes from original relation of derivation distribution
function of hot electrons $(k_\parallel u_\perp \partial f / \partial
u_\parallel) / \gamma $ (\citet{1986ApJ...307..808W}, Equation A8). As shown below, this term is negligible, and thus in our computations we use
$G^{\pm} = g_1$.

The sum $h^+ + h^-$ expresses a simultaneous effect of the
both operators $h^+, h^-$ on one upper-hybrid wave described by the specific
k-vector. In our computations we used both operators, but we found that term $h^-$ was always at least two orders lower than $h^+$ and the typical difference was ten orders.

 Equations \ref{eq1}-\ref{eq5} are valid for $k_\perp \gg k_\parallel$, see
\citet{1986ApJ...307..808W}. We note that comparing these relations with those in
the paper by \citet{2004SoPh..219..289Y} the growth rate (relation~\ref{eq2},
$G^{\pm}=g_1$) is proportional to $s^3$, not to $s^2$.

The integration is done for velocities up to their maxima
\begin{equation}\label{eq7}
u_{\perp, \mathrm{max}} = \frac{\sqrt{-1 + (1+r_\mathrm{pB}^2 - s^2) v_\mathrm{t}^2 \sigma^2 + 3 v_\mathrm{tb}^2 (1 + \lambda \sigma^2)}}{ \sqrt{1- (1+r_\mathrm{pB}^2) v_\mathrm{t}^2 \sigma^2 - 3 v_\mathrm{tb}^2 (1 + \lambda \sigma^2)}},
\end{equation}
from which we get the followning conditions for maximal and minimal value of $\lambda$ and $\sigma$ (expressions under root in previous equations equal zero):
\begin{eqnarray}
    \label{eq8}
    \sigma_\mathrm{max}(v_\mathrm{t}, v_\mathrm{tb}, s, \lambda) &=& \frac{\sqrt{1 - 3 v_\mathrm{t}^2}}{\sqrt{(1+r_\mathrm{pB}^2 - s^2) v_\mathrm{t}^2 + 3 v_\mathrm{tb}^2 \lambda}}, \\
    \label{eq9}
    \sigma_\mathrm{min}(v_\mathrm{t}, v_\mathrm{tb}, s, \lambda) &=& \frac{\sqrt{1 - 3 v_\mathrm{t}^2}}{\sqrt{(1+r_\mathrm{pB}^2) v_\mathrm{t}^2 + 3 v_\mathrm{tb}^2 \lambda}}, \\
    \label{eq10}
    \lambda_\mathrm{min}(v_\mathrm{t}, v_\mathrm{tb}, s) &=& \frac{(-1 - r_\mathrm{pB}^2 + s^2)v_\mathrm{t}^2}{3 v_\mathrm{tb}^2}.
\end{eqnarray}

For the comparison below, we add values of $\sigma$ and $\lambda $ for the
maximal growth rate as follows from analytical estimations made by
\citet{1986ApJ...307..808W}
\begin{eqnarray}
    \label{eq11}
    \sigma_{\Gamma_\mathrm{max}} &=& \frac{1}{\sqrt{2(r_\mathrm{pB}^2 + 1) }v_\mathrm{t}^2}, \\
    \label{eq12}
    \lambda_{\Gamma_\mathrm{max}} &=& \frac{s^2}{2},
\end{eqnarray}
when we suppose $u_{\perp,\mathrm{max}} = \sqrt{2}v_\mathrm{t}$.

\subsection{Methods}
We used Python SympPy\footnote{http://www.sympy.org} library for analytical
application and Python SciPy\footnote{http://www.scipy.org} library for
numerical computations of growth rate equations
\ref{eq1}-\ref{eq5},\ref{eq7}-\ref{eq10}.

Generally, the analytical expressions for the growth rate (relations
\ref{eq1}-\ref{eq5}) depend on $v_\mathrm{t}$, $v_\mathrm{tb}$, $\lambda$,
$\sigma$, $s$ and $r_\mathrm{pB}$. Their numerical solutions are made in
several steps as described below.

First of all is the choice of thermal velocities $v_\mathrm{t}$, $v_\mathrm{tb}$.
Then we selected the $r_\mathrm{pB}$ interval in which computations are made
and steps in this interval. We usually use the interval $r_\mathrm{pB} =
3-20$ with the step as $\Delta r_\mathrm{pB} = 0.03$.

For each value of $r_\mathrm{pB}$ we searched for $s$ which fulfill
$s=s_0-m, \ldots ,s_0+n$, where $s_0 = \mathrm{Round}(r_\mathrm{pB})$ is
the round value of $r_\mathrm{pB}$. Numbers $m,n \in \mathbb{N}$
creates the interval of $s$ for which the growth rate can be computed.
However, we limited this interval only for $s$, maximal values of
the growth rate in this interval are at least one hundredth of the maximal growth rate
for $s_0$.

Next, values of $v_\mathrm{t}$, $v_\mathrm{tb}$, $r_\mathrm{pB}$ and $s$ were
chosen, and values of remaining variables $\lambda$ and $\sigma$, which
correspond to the k-vector components of the upper-hybrid waves, need to be
specified. We chose these values in the intervals $\sigma \in
(\sigma_\mathrm{min}, \sigma_\mathrm{max})$ and $\lambda \in
(\lambda_\mathrm{min}, \lambda_\mathrm{max})$ on equidistant lattice.  If
the maximal values of $\sigma_\mathrm{max}$ and $\lambda_\mathrm{max}$ are
too large or not defined, the upper boundaries are chosen as ten times of
their minimal values. It was found to be sufficient in all cases.

Only then it is possible to compute the growth rate given by relations between \ref{eq1}-\ref{eq5}
for parameters $v_\mathrm{t}$, $v_\mathrm{tb}$ and $r_\mathrm{pB}$ for each
$s$ and in all points of the map $\lambda - \sigma$. In following step we summed
the maps in each specific ($\lambda - \sigma$)-point over $s$. Finally, in
the resulting map we searched for the point ($\lambda_{\Gamma
max},\sigma_{\Gamma max}$), where is the highest value of the growth rate,
which is then that determined for chosen $r_\mathrm{pB}$. The wave number
components of the upper-hybrid wave with the maximal growth rate is then
determined by Equation \ref{eq6}.

\section{Results}

We solved the above described equations for the parameters shown in
Table~\ref{tab1}, for $r_\mathrm{pB}=3-20$ and $B=100$~G.  In Models 1-3, we
changed background temperature while temperature of the hot electrons remains
constant. Models 4 and 5 consider higher temperatures of the hot electrons
comparing with Model 1, while the temperature of the background plasma is
fixed.

\begin{table}[h!]
\caption{Computation parameters.} \label{tab1} \centering
\begin{tabular}{ccr}
\hline\hline
Model No. & $v_\mathrm{t}$ & $v_\mathrm{tb}$ \\
\hline
1 & 0.1~c & $0.018$~c \\
2 & 0.1~c & $0.009$~c \\
3 & 0.1~c & $0$~c \\
4 & 0.2~c & $0.018$~c \\
5 & 0.3~c & $0.018$~c \\
\hline
\end{tabular}
\end{table}

\begin{figure}[h!]
\centering
\includegraphics[width = 0.49\textwidth]{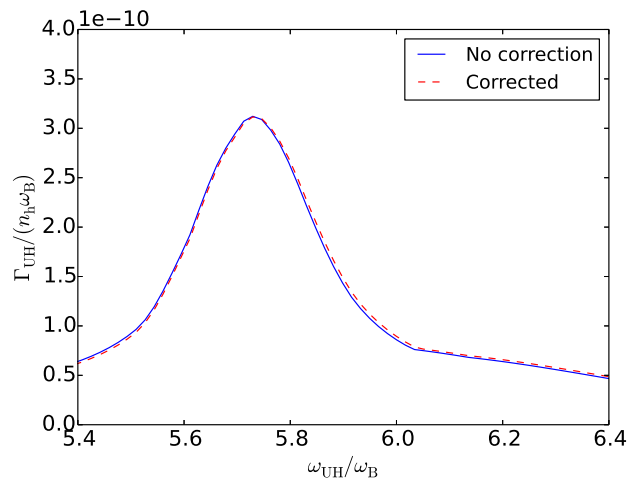}
\caption{Example of the comparison of the growth rates computed with and without the term
$g_2^\pm$ (Equation \ref{eq2}), for Model 1 and $s=6$. We note that the growth rate at this and following figures
are normalized by the term $n_\mathrm{h} \omega_\mathrm{B}$.} \label{Fig8}
\end{figure}

First, we tested an importance of the correction term $g_2^\pm$ in
Equation~\ref{eq6a} for computation of the growth rate. An example of such
comparison  is shown in Figure \ref{Fig8}. We found that in all models
differences in values of the growth rate in models with and without this term
is less than 5~\%, and at their maxima even smaller. Thus, we consider the
effect of this term negligible and in all following computations we suppose
$g_2^\pm = 0$.

\begin{figure}[!h]
\centering
\includegraphics[width = 0.45\textwidth]{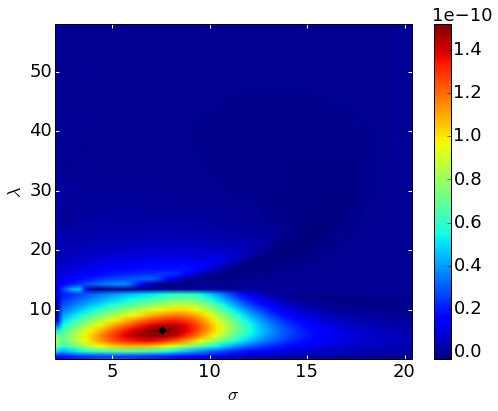}\\
\includegraphics[width = 0.45\textwidth]{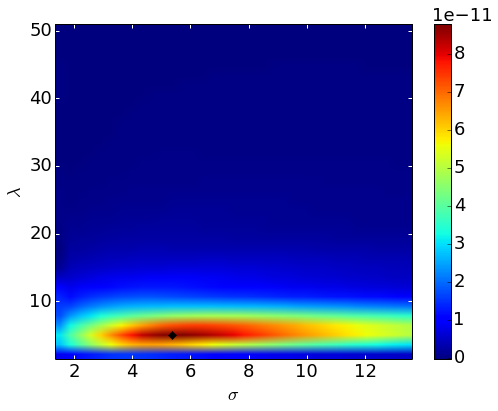}\\
\includegraphics[width = 0.45\textwidth]{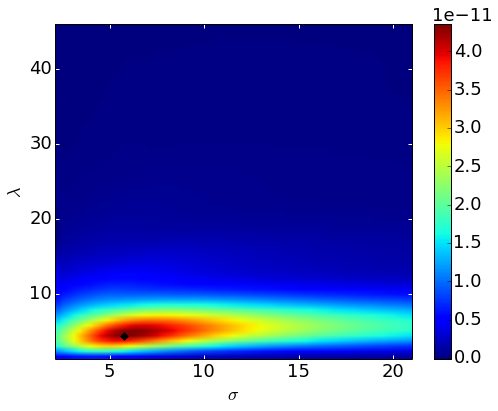}
\caption{Growth rates  in the $\lambda - \sigma$ space: Model 1 (top), Model 4 (middle),
Model 5 (bottom); all computed for $s = 6$ and $r_\mathrm{pb}$, for which the growth rate is maximal.
In Model 1 $r_\mathrm{pb} = 5.8$, compare with Figure~\ref{Fig5}.
Black diamonds show positions of maximal values of the growth rate.}
\label{Fig2}
\end{figure}

\begin{figure}[!h]
\centering
\includegraphics[width = 0.49\textwidth]{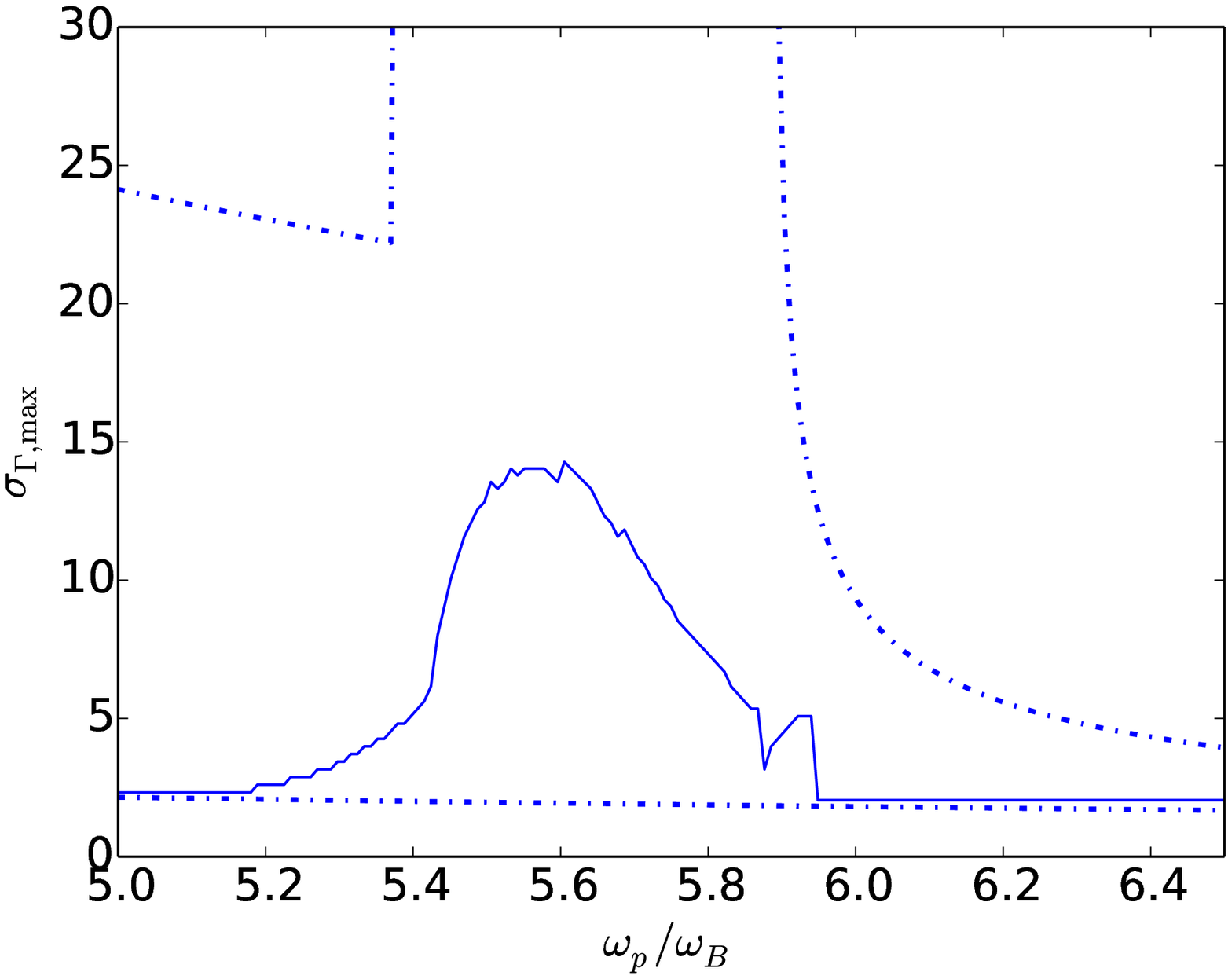}\\
\includegraphics[width = 0.49\textwidth]{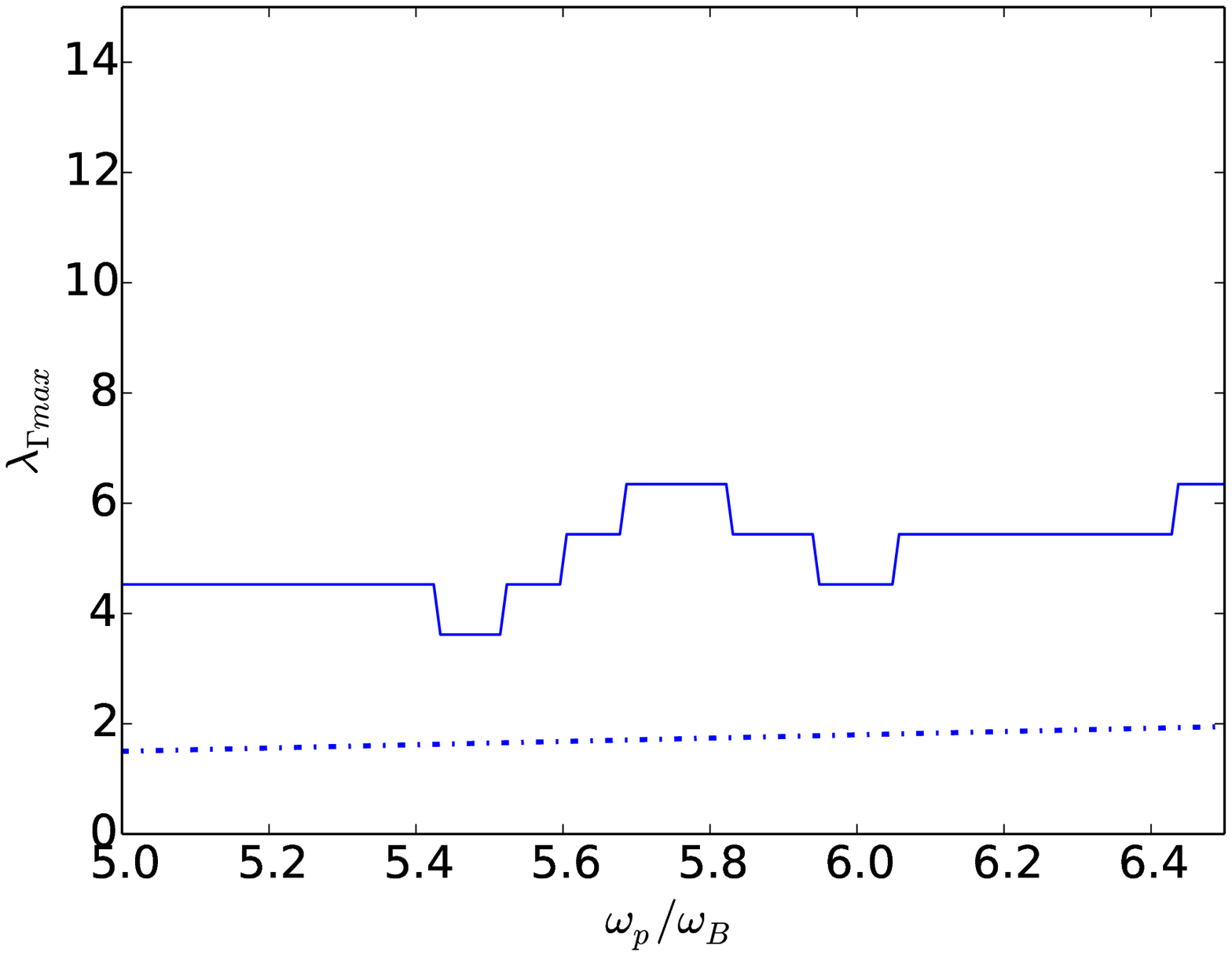}
\caption{Values of $\sigma$ (top) and $\lambda$ (bottom) for Model 1 and $s=6$, where the growth rate
has the maximal value depending on $r_\mathrm{pB}$. Solid lines show values of the maximal
growth rate, dash-dotted lines are boundaries of the $\sigma - \lambda$ space. Where dash-dotted lines are not shown,
the boundaries are outside the presented region.} \label{Fig5}
\end{figure}

In computations of the growth rate in all Models (see Table \ref{tab1}) we
followed steps described in Methods. Thus, we obtained many maps in the
$\lambda - \sigma$ space, corresponding to the k-vector space. We note that the
relations between the $\lambda - \sigma$ and k-vector space is given by relation \ref{eq6}.

\begin{figure}[h!]
\centering
\includegraphics[width = 0.49\textwidth]{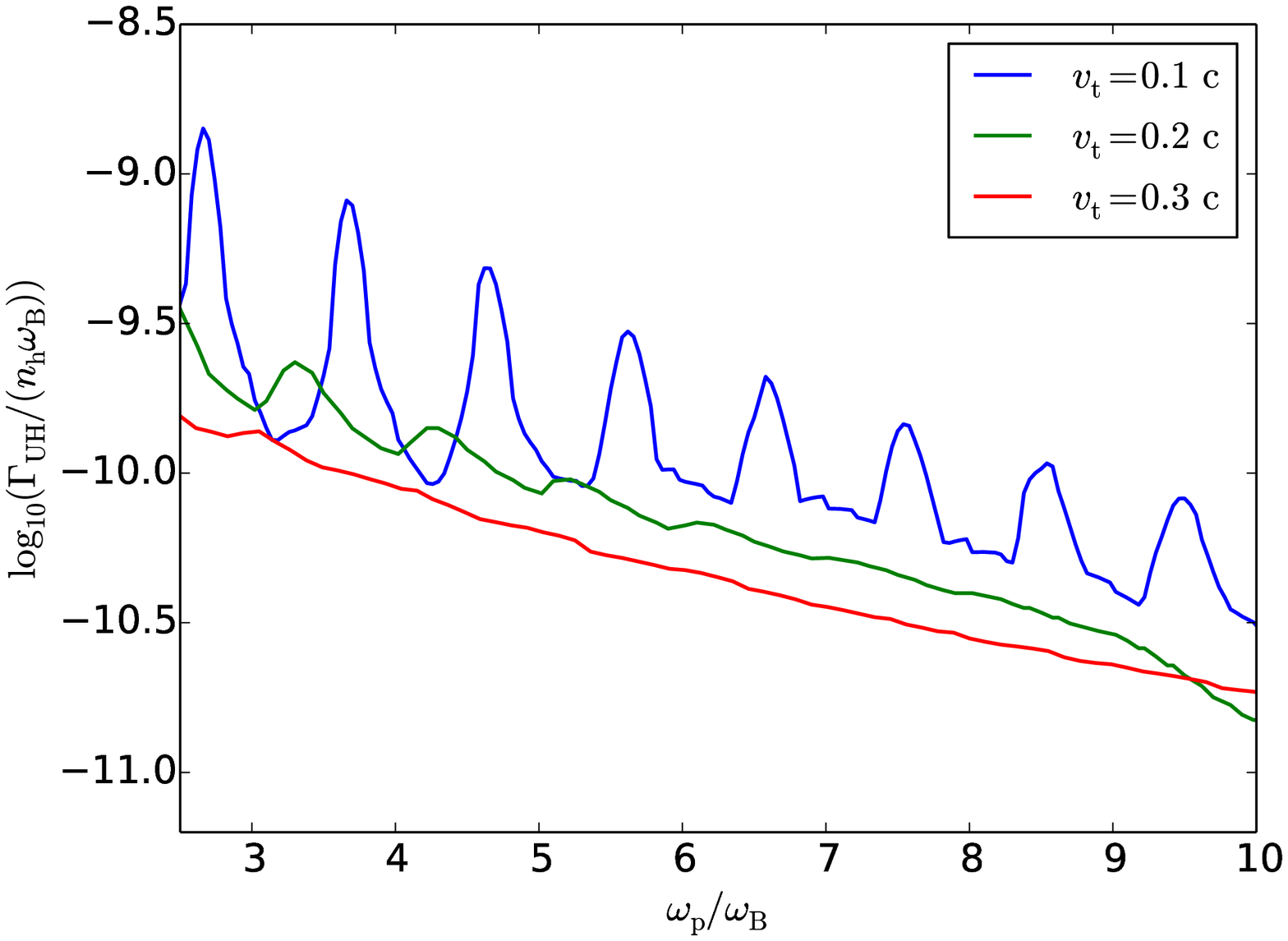} \\
\includegraphics[width = 0.49\textwidth]{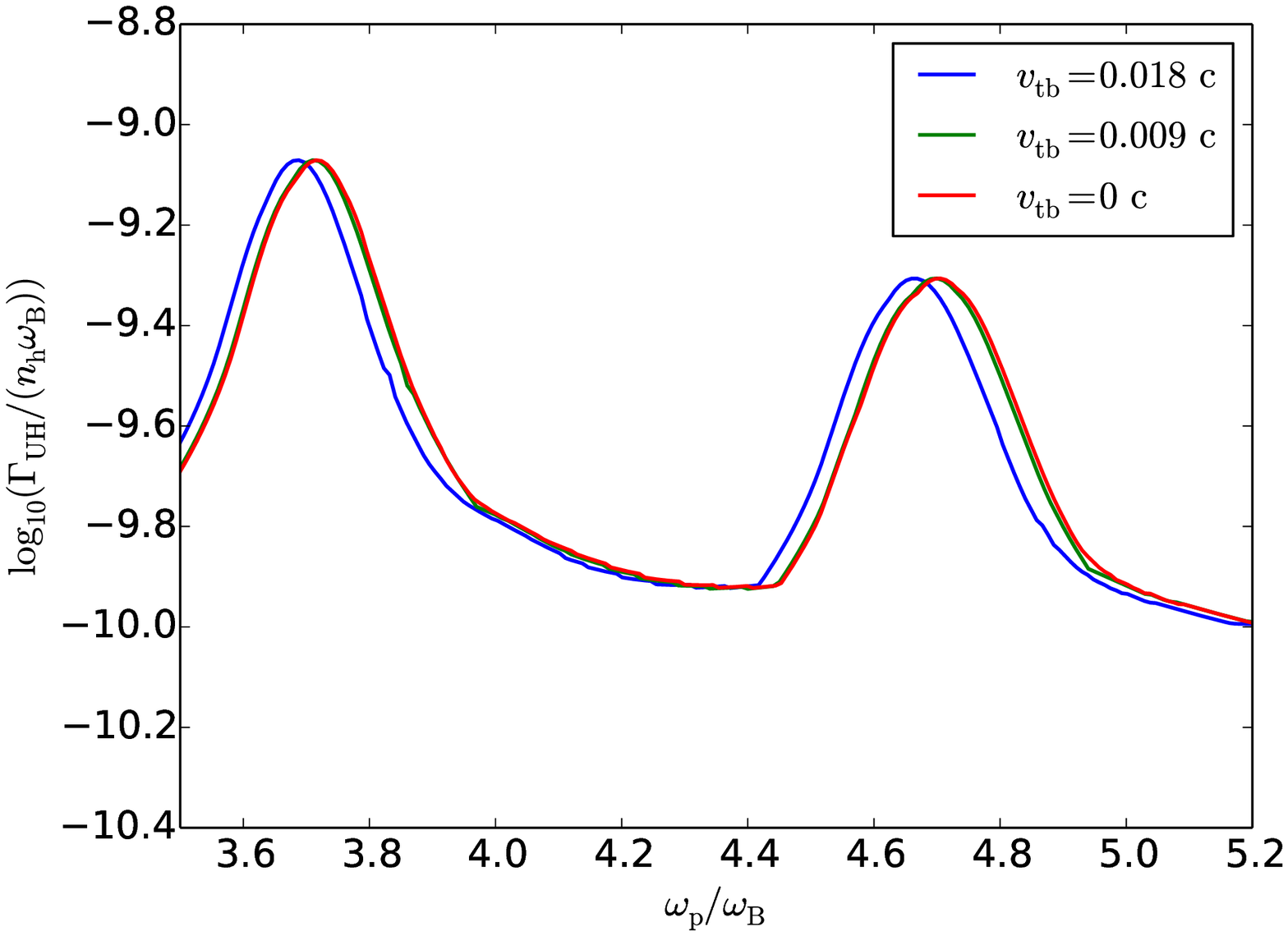}
\caption{Growth rates in Models shown in Table~\ref{tab1} in dependence on
$r_\mathrm{pB} = \omega_\mathrm{p}/\omega_\mathrm{B}$. {\it Upper part:} Growth
rates for Models 1, 4 and 5 with the fixed background temperature
$v_\mathrm{tb} = 0.018$~c. {\it Bottom part:} Growth rates for Models 1-3 with
the fixed temperature of hot electrons $v_\mathrm{t} = 0.1$~c; detailed view.} \label{Fig1}
\end{figure}

Examples of such maps for $s = 6$ at their maxima are shown in Figure
\ref{Fig2}. In each of these maps the maximal growth rate is indicated. Then we summed the maps over $s$ and in the resulting map we
searched for the maximal growth rate. This growth rate then corresponds to one
point of the curve in Figure \ref{Fig1}.

To compare the computed and analytically estimated (relations
\ref{eq11}-\ref{eq12}) k-vectors, which correspond to the maximal growth
rate for the cases presented in Figure~\ref{Fig2}, these are shown in Table
\ref{tab2}. Thus, the assumption ($k_\perp \gg k_\parallel$) used in
derivation of the growth-rate relations was fulfilled in all studied Models.
All computed k-vectors are smaller than that analytically predicted. Their
computed perpendicular components are systematically two times less than analytical
ones. This is due to assumptions  and simplifications made in analytical
estimations of  these values. Computed values of $k_\parallel$ decrease with
increasing temperature in contradiction with analytical predictions.

Examples of values $\sigma_{\Gamma, \mathrm{max}}, \lambda_{\Gamma,
\mathrm{max}}$ for $v_\mathrm{t}=0.1$~c (Model 1), where the growth rate has
the maximal value depending on $r_\mathrm{pb}$, are presented in Figure
\ref{Fig5}.
 Values of $\sigma_{\Gamma, \mathrm{max}}$ for $r_\mathrm{pB} \sim 5$ and $r_\mathrm{pB}$ > $6$ are close to the lower boundary
 (given by Equation \ref{eq9}). However, in the region around the growth-rate maximum they deviate from the lower boundary.
 Values of $\lambda_{\Gamma, \mathrm{max}}$ are changing only slightly.
 The steps in curves are due to discretisation of the $\lambda - \sigma$ space.
 Note that the value $\lambda_{\Gamma, \mathrm{max}} \approx 5$ (Figure \ref{Fig5}, bottom)
 is less then predicted value $s^2 /2 = 18$.

Based on many such $\lambda - \sigma$ maps the growth-rate in dependence on the
$r_\mathrm{pB} = \omega_\mathrm{p}/\omega_\mathrm{B}$ were computed for all
Models according to Table~\ref{tab1}. These growth rates are shown in Figure
\ref{Fig1}. The upper part of this Figure shows an effect of the temperature
increase of the hot electrons for fixed temperature of the background plasma
electrons, and its bottom part an effect of the temperature increase of the
background plasma electrons for fixed temperature of hot electrons.

As seen in the upper part of this Figure, Model 1 with the lower velocity of
hot electrons have higher growth rates than those in Models 4 and 5 for all
$r_\mathrm{pB} = \omega_\mathrm{p}/\omega_\mathrm{B}$.
Furthermore, the temperature increase of hot electrons smooths
peak maxima in the growth rate. While Model 1 shows distinct maxima, in Model 5
the maxima for $s>5$ are hardly recognisable.

On the other hand, the change of the thermal velocity of background electrons
in Models 1-3 for fixed temperature of hot electrons (see the bottom part of
the Figure \ref{Fig1}) do not change values of the growth rate, but the
temperature increase shifts the maxima slightly to lower $r_\mathrm{pB}$. The
values of maxima generally decrease with an increase of $r_\mathrm{pB}$.

\begin{table}[h!]
\caption{Positions of maximal growth rates (diamonds) from Figure~\ref{Fig2} in
$\vec{k}$-space for $s=6$. All values agrees $k_\parallel \ll k_\perp$.
Computed values are lower than analytically predicted.} \label{tab2} \centering
\begin{tabular}{ccccc}
\hline\hline
Model no. & $k_\parallel$ & $k_{\parallel,\mathrm{theor}}$ &  $k_\perp$ & $k_{\perp,\mathrm{theor}}$ \\
~ & [cm$^{-1}$] & [cm$^{-1}$] & [cm$^{-1}$] & [cm$^{-1}$] \\
\hline
1 & 0.078 & 0.049 & 1.49 & 2.48 \\
4 & 0.054 & 0.086 & 0.66 & 1.24 \\
5 & 0.034 & 0.117 & 0.41 & 0.83 \\
\hline
\end{tabular}
\end{table}

\begin{figure}[h!]
\centering
\includegraphics[width = 0.45\textwidth]{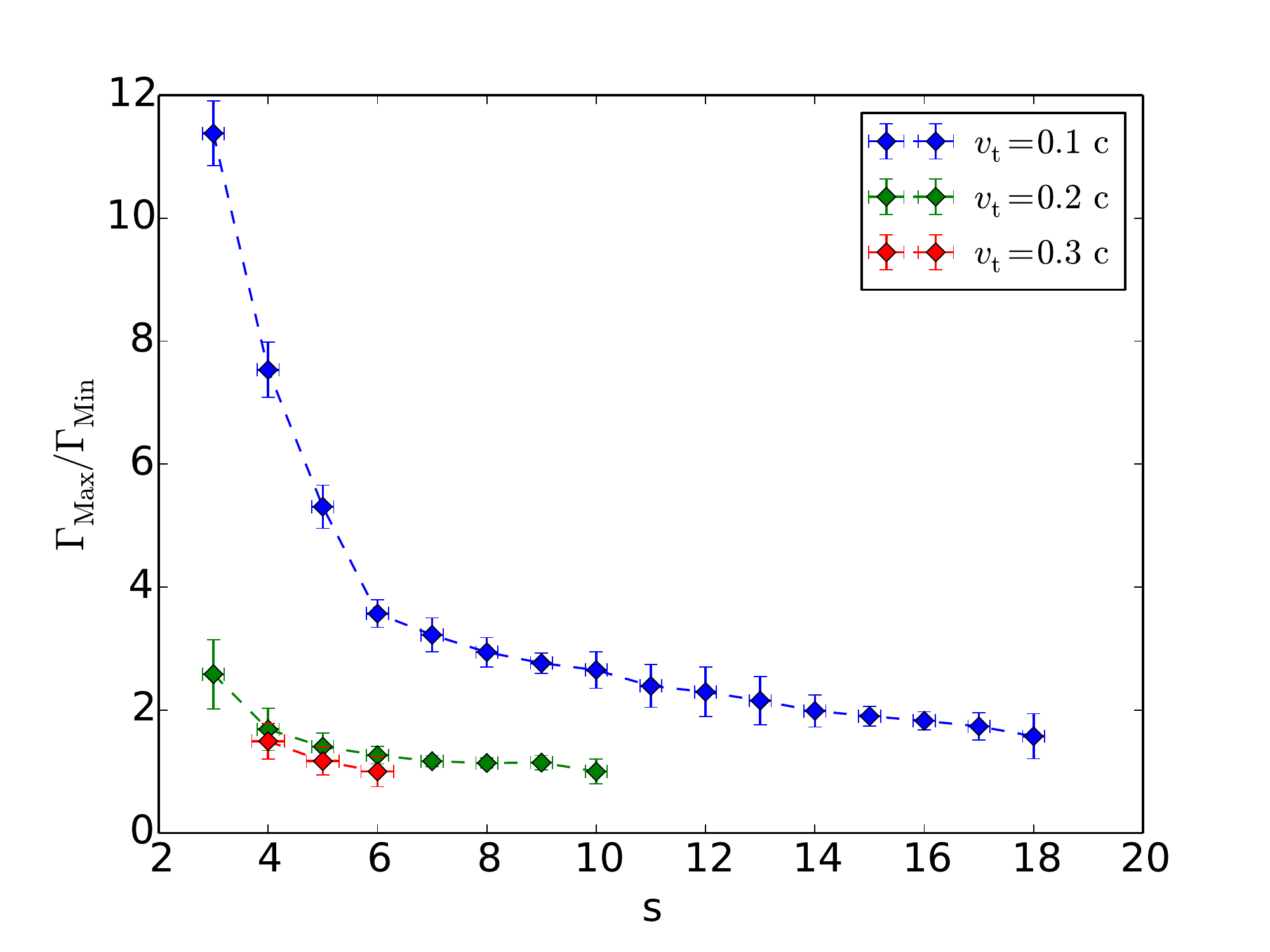}
\caption{Ratio between maxima and neighbouring minima of growth rate depending on $s$ for Models 1, 4 and 5. } \label{Fig6}
\end{figure}

\begin{figure}[h!]
\centering
\includegraphics[width = 0.45\textwidth]{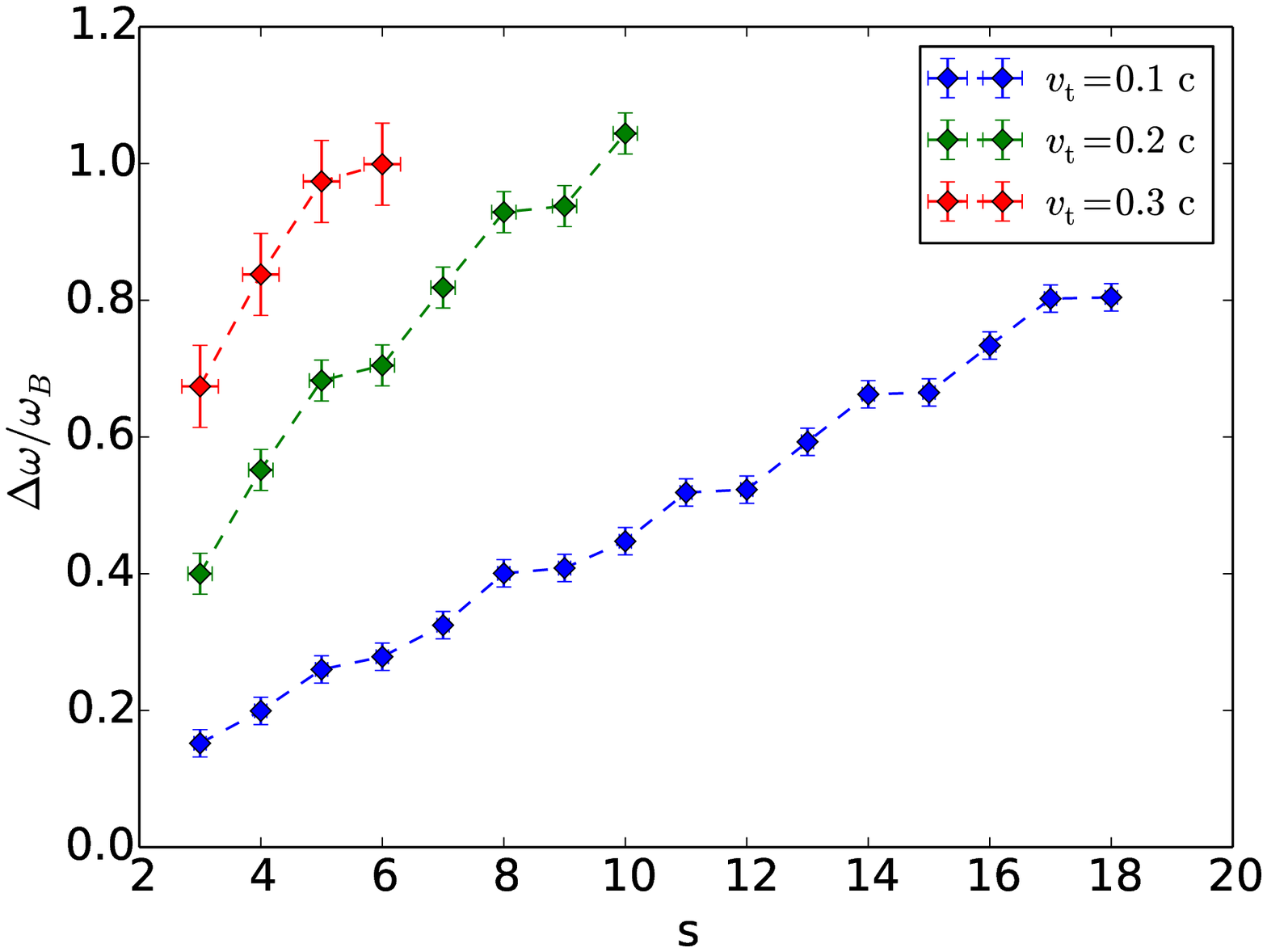}\\
\includegraphics[width = 0.45\textwidth]{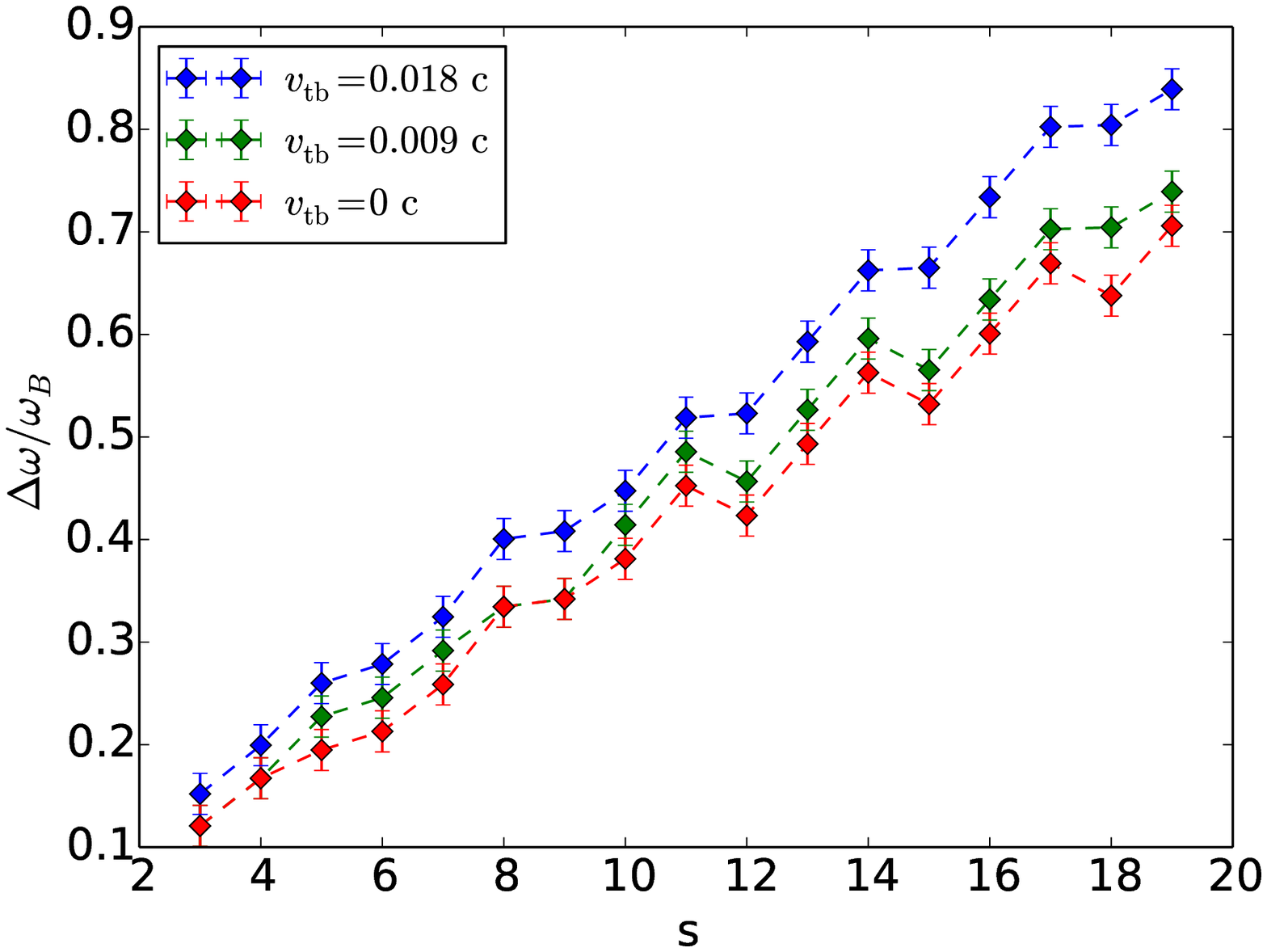}
\caption{Relative frequency shift $\Delta \omega$ of growth-rate maxima between predicted values by $\omega_\mathrm{UH} = s \omega_\mathrm{B}$ and its computed position (see Figure~\ref{Fig1}). Top Models 1, 4 and 5 with changing hot electrons temperature; bottom Models 1-3 with changing background temperature.} \label{Fig7} 
\end{figure}

As already mentioned, maxima of the growth rates are more distinct for lower
temperatures of the hot electrons. Therefore in Figure~\ref{Fig6} we present
ratios of these maxima and neighbouring minima of the growth rate in dependence
on $s$. We note that there is a simple relation between $s$ and $r_\mathrm{pB}$; $s
 = \sqrt{r_\mathrm{pB}^2 +1}$. As seen in Figure~\ref{Fig6}, the ratios decrease with increasing $s$. It
does not depend on the temperature of background electrons.

There is another interesting effect of temperatures of the both hot and
background plasma electrons. Namely, by increasing these temperatures, the growth-rate maxima are shifted to lower frequencies as shown in Figure \ref{Fig7}.

\begin{figure}[h!]
\centering
\includegraphics[width = 0.49\textwidth]{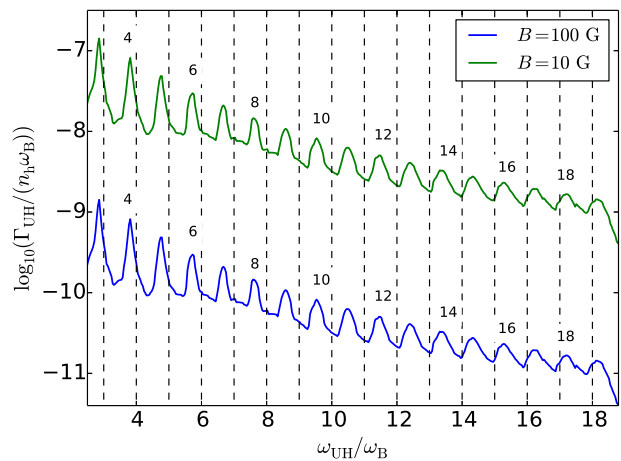}
\caption{Growth rate for Model 1 (B= 100 G, $f_B = \omega_\mathrm{B}/2\pi =$ 280 MHz) (blue line) and
the model with similar parameters to Model 1, except the
magnetic field intensity $B = 10$ G ($f_\mathrm{B} = \omega_\mathrm{B}/2\pi =$ 28 MHz) in dependence on $\omega_\mathrm{UH}/\omega_\mathrm{B}$.
Numbers close the growth-rate maxima designate $s$ from computations.
We note that the growth rate at this figure is normalized by the term $n_\mathrm{h} \omega_\mathrm{B}$.} \label{Fig9}
\end{figure}

In Figure 7 we present this effect in a way that is more appropriate for zebra pattern analysis. Here we show growth-rate
dependence on the ratio
$\omega_\mathrm{UH}/\omega_\mathrm{B} = s$ for Model 1, meaning that, with the
magnetic field intensity $B = 100 $ G ($f_\mathrm{B} = \omega_\mathrm{B}/2\pi
=$ 280 MHz) and for the model with similar parameters as in Model 1, except
that the magnetic field intensity is $B = 10$ G ($f_\mathrm{B} =
\omega_\mathrm{B}/2\pi =$ 28 MHz). At the horizontal axis of this
figure we used the ratio $\omega_\mathrm{UH}/\omega_\mathrm{B}$, because
$\omega_\mathrm{UH}$ can be directly determined from frequencies of the
observed zebra stripes. As shown here, the growth-rate maxima do not correspond
to a simple resonance conditions (relation~\ref{eq00}), which is used in
estimations of the magnetic field and plasma density in the zebra radio sources
from observed zebras. The maxima are shifted to lower frequencies, and this
shift increases with the increase of $s$. Moreover, as seen in
Figure~\ref{Fig9}, the relative bandwidth of the growth-rate maxima also increases with
the increase of $s$.

\begin{figure}[h!]
\centering
\includegraphics[width = 0.49\textwidth]{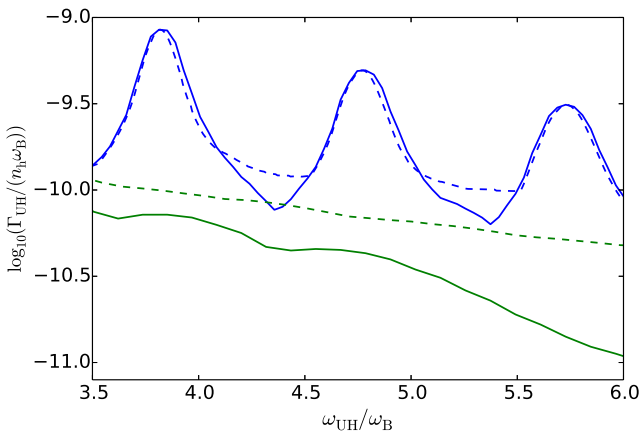}
\caption{Comparison of the growth rates with summation over $s$ and for $s$ with the maximal growth rate,
for Model 1 (blue line) and Model 5 (green line).
Dashed lines show summation over $s$ and solid lines correspond to $s$ with the maximal growth rate.} \label{Fig10}
\end{figure}

We also analyzed an effect of summation over $s$ in
relation~\ref{eq1} in comparison with the case for $s$ with the maximal growth
rate, see Figure~\ref{Fig10} made for Models 1 and 5. As seen here, for lower
temperatures $v_\mathrm{t}$ the summation increases values of the growth-rate
minima and for higher temperatures the growth rate is enhanced in the whole
range. We found that if for a given $s$ the growth rate decreases steeply from
its maximum (the case with lower $v_\mathrm{t}$), the summation is important at
places where the individual orders overlap. But if the decrease of the growth
rate from its maximum is gradual (the case with higher $v_\mathrm{t}$), the
summation influences the growth rate in a broad range of $r_\mathrm{pB}$.

\section{Discussion}
In the present paper we studied effects of non-zero temperatures of the both
hot and background plasma on the growth rate.
We showed that including effects of non-zero temperatures leads to shifts
in the growth-rate maxima to lower frequencies, comparing frequencies derived
from the simple resonance condition $\omega_\mathrm{UH} = s \omega_\mathrm{B}$.
We found that these shifts are greater for higher values of the cyclotron
harmonic number $s$. Because the simple resonance condition is used in
estimations of the magnetic field and plasma density in the zebra radio
sources, these shifts influence such estimations.

Therefore, let us estimate an error in the magnetic field determination from
some zebra stripes using the simple resonance condition. For this purpose, let
us assume the zebra stripe on the frequency $f = f_\mathrm{UH} =
\omega_\mathrm{UH}/2\pi =$ 504 MHz (assuming the emission on the fundamental
frequency) with $s =$ 18. If we use the simple resonance condition then for
such high $s$ the ratio of $r_\mathrm{pB} =
\omega_\mathrm{p}/\omega_\mathrm{B}$ $\approx s$, and thus $f_B =
\omega_\mathrm{B}/2\pi =$ 28 MHz, which gives the magnetic field $B =$ 10 G,
see also Figure~\ref{Fig9}. But, for modified Model~1 with $B =$ 10 G we
found that the emission maximum for $s =$ 18 is shifted  to
$\omega_\mathrm{UH}/\omega_\mathrm{B} =$ 17.3. From observations, we have
$f_\mathrm{UH} = \omega_\mathrm{UH}/2\pi =$ 504 MHz, thus $f_\mathrm{B} =
504/17.3 \approx$ 29.1 MHz, which gives the magnetic field $B =$ 10.4 G. This
example shows a 4~\% error in magnetic field estimation. As shown in
Figure~\ref{Fig7}, the shift of the growth rates grows linearly with $s$.
However, the relative error of determining the magnetic field does not
depend on $s$. For higher temperatures of hot electrons errors in magnetic
field estimations are larger. For example, for Model 4 (recognisable
maxima only for $s<$ 11) this error is $\approx$16~\%.

The growth rate is proportional to the density of the hot plasma $n_\mathrm{h}$
and inversely proportional to the magnetic field. If we multiply the growth
rate for Model 3 (with zero temperature of the background plasma), expressed in
our paper as $\Gamma_\mathrm{UH}/(\omega_\mathrm{B} n_\mathrm{h})$ by the
density $n_\mathrm{h} = 10^8~\mathrm{cm}^{-3}$ we obtain the growth rate, which
agrees with that by \citep{1986ApJ...307..808W}.

In agreement with the results by \citet{2004SoPh..219..289Y} we found that for
a relatively low temperature of the hot electrons (Model 1) the dependence of
the growth rate vs. the ratio of the electron plasma and electron
cyclotron frequencies expresses distinct peaks and increasing this temperature
(Models 4 and 5) these peaks are smoothed, especially for high $s$.
This growth-rate behaviour differs from those in previous studies
\citep{1975SoPh...44..461Z,1986ApJ...307..808W}. The difference is caused by
relativistic corrections used in the present paper in both resonance terms in
Equation \ref{eq13} and in selection of maximal growth rates in \vec{k}-maps.
We note that a similar effect in growth-rate behaviour was found by
\citet{2007SoPh..241..127K}. Namely, considering the power-law and loss-cone
distribution function of hot electrons, they showed that stripes of a zebra
pattern become more pronounced with an increase of the loss-cone opening angle
and the power-law spectral index.

We also studied effects of the neglected term in the growth rate with
$\frac{\partial f}{\partial u_\parallel}$ (\citet{1986ApJ...307..808W} Equation
A8). We found that these effects are negligible.

\section{Conclusions}

We found that the growth-rate maxima of the upper-hybrid waves for non-zero
temperatures of the both hot and background plasma are shifted towards lower
frequencies comparing to the zero temperature case.
It was found that this shift increases with an increase of the harmonic number
$s$ of the electron cyclotron frequency and temperatures of the both hot and
background plasma components.

We showed that this shift of the growth-rate maxima influences estimations of
the magnetic field strength in sources of observed zebras.
In agreement with previous studies we found that for a relatively low
temperature of the hot electrons the dependence of the growth rate vs.
the ratio of the electron plasma and electron cyclotron frequencies expresses
distinct peaks and increasing this temperature these peaks are smoothed.

We found that in some cases the values of components of the wave number vector
of the upper-hybrid wave for the maximal growth rate strongly deviate from
their analytical  estimations.
Validity of the assumptions used in derivation of the model equations was
confirmed.

\begin{acknowledgements}
The authors thank G.P. Chernov for useful comments. We acknowledge
support from Grants P209/12/0103. Access to computing and storage facilities
owned by parties and projects contributing to the National Grid Infrastructure
MetaCentrum provided under the programme "Projects of Projects of Large
Research, Development, and Innovations Infrastructures" (CESNET LM2015042), is
greatly appreciated. L.Y. acknowledges support by the RFBR grant (No. 16-02-00254).
\end{acknowledgements}

\bibliographystyle{aa}
\bibliography{maser}

\end{document}